\documentstyle[12pt,epsf]{article}
\newcommand{\be}{\begin{equation}}
\newcommand{\ee}{\end{equation}}
\newcommand{\ba}{\begin{eqnarray}}
\newcommand{\ea}{\end{eqnarray}}
\newcommand{\bi}{\begin{itemize}}
\newcommand{\ei}{\end{itemize}}

\def\lsi{\raise0.3ex
\hbox{$<$\kern-0.75em\raise-1.1ex\hbox{$\sim$}}}
\def\gsi{\raise0.3ex
\hbox{$>$\kern-0.75em\raise-1.1ex\hbox{$\sim$}}}
\newcommand{\lsim}{\mathop{\lsi}}

\begin{document}

\begin{titlepage}
\begin{flushright}
JYFL-1/00\\
hep-ph/0002008\\
%hep-ph/0002008
\end{flushright}
\begin{centering}
\vfill

{\bf PRODUCTION OF TRANSVERSE ENERGY FROM MINIJETS
IN NEXT-TO-LEADING ORDER PERTURBATIVE QCD}

\vspace{0.5cm}
 K.J. Eskola\footnote{kari.eskola@phys.jyu.fi}
and K. Tuominen\footnote{kimmo.tuominen@phys.jyu.fi}

\vspace{1cm}
{\em  Department of Physics, P.O.Box 35, FIN-40351
Jyv\"askyl\"a, Finland\\}

\vspace{1cm}
{\bf Abstract}

\end{centering}

\vspace{0.3cm}\noindent
We compute in next-to-leading order (NLO) perturbative QCD the transverse
energy carried into the central rapidity unit of hadron or nuclear
collisions by the partons freed in the few-GeV subcollisions. The
formulation is based on a rapidity window and a measurement function
of a new type. The behaviour of the NLO results as a function of the
minimum transverse momentum and as a function of the scale choice is
studied. The NLO results are found to be stable relative to the leading-order 
ones even in the few-GeV domain.
\vfill 

\end{titlepage}

Below the transverse momenta of observable jets in hadronic
collisions, $p_T\lsim 5$ GeV, but still within the applicability
domain of perturbative QCD, $\Lambda_{\rm QCD}\ll$~$p_T$, there is a
region of semi-hard parton production, $p_T$~$\sim$~1..2~GeV. In
ultrarelativistic heavy ion collisions this region is especially
important in the formation of Quark Gluon Plasma at the future
colliders BNL-RHIC and CERN-LHC/ALICE: 
at high collision energies, the few-GeV QCD-quanta,
minijets, dominate the initial transverse energy production
\cite{BM87,KLL87,EKL89} at central rapidities 
during the first fractions of fm/$c$.  The initial energy densities
for further evolution of the system can thus be estimated
based on perturbative QCD (pQCD)\cite{EKRT99}.

As work $pdV$ is done during the expansion of the system, the initial
transverse energy is not, however, directly measurable in $AA$
collisions: the final state $E_T$ in the central rapidity unit has
been estimated to be only $1/6$ ($1/3$) of the initially produced
$E_T$ at the LHC (RHIC) \cite{EKRT99}. The final $E_T$ depends
practically linearly on the initial one, so for making reliable
estimates of the measurable $E_T$ the initial $E_T$ needs to 
be computed as accurately as possible.

The average initial transverse energy produced perturbatively in the
central rapidity unit $\Delta Y$ of an $AA$ collision at an impact
parameter ${\bf b}$ can be computed as \cite{EKL89}
\begin{equation}
\overline E_T^{AA}({\bf b,}\sqrt s,p_0,\Delta Y) 
= T_{AA}({\bf b})\sigma\langle E_T\rangle_{p_0, \Delta Y}.
\end{equation}
The standard nuclear overlap function $T_{AA}({\bf b})$ accounts for
the nuclear collision geometry ($T_{AA}(0)\approx A^2/(\pi R_A^2)$ for
$A\sim200$ \cite{EKL89}).  The first moment of the semi-inclusive
$E_T$ distribution, $\sigma\langle E_T\rangle_{p_0, \Delta Y}$, is the
pQCD quantity we formulate and compute in NLO below. The scale $p_0$
is the smallest transverse momentum scale in the
computation. It also governs the formation time of the system through
$\tau_0\sim 1/p_0$. To compute the actual values of initial $E_T$,
$p_0$ has to be determined dynamically by introducing additional
(nonperturbative) phenomenology, see e.g.  \cite{EKRT99}.  We
emphasize that in this work we do not discuss this but aim to show
that the NLO pQCD formulation is field-theoretically well defined and
perform a rigorous pQCD computation of $\sigma\langle E_T\rangle_{p_0,
\Delta Y}$. Therefore, at this level $p_0$ is a fixed external
parameter with no other physical significance than $p_0\gg\Lambda_{QCD}$.  
The dependence of $\sigma\langle E_T\rangle$ on $p_0$ will be explicitly 
studied and the rigorous NLO computation presented here then sets 
the stage for more phenomenological analyses.

We generalize the LO formulation \cite{EKL89} of $\sigma\langle E_T\rangle$
to NLO  by introducing a new type of infrared (IR) safe
measurement functions \cite{KS92} which contain both the rapidity
acceptance and the definition of perturbative collisions.  In getting
from the $4-2\varepsilon$ dimensional squared matrix elements to the
physical quantities we apply the procedure by S. Ellis, Kunszt and
Soper (EKS) \cite{KS92,EKS89,EKS90,KUNSZT}.  

The new results obtained in our study can be summarized as follows: A
consistent and well-defined NLO pQCD formulation of $\sigma\langle
E_T\rangle_{p_0, \Delta Y}$ exists. Eventhough towards the few-GeV
region the NLO results could grow rapidly relative to the LO, a stable
behaviour of the NLO results is discovered in the range $1...2\,{\rm
GeV}\lsim p_0\lsim10\,{\rm GeV}$. This signals of the applicability of
the pQCD in the semihard domain.  The NLO computation also brings in a new
kinematical region not present in the LO, thus increasing the amount 
of perturbatively computable $E_T$ in nuclear and hadronic collisions.  
Previously, as the actual NLO contributions were not known, various 
{\it ad hoc} $K$-factors to the LO formulation \cite{EKL89} have been 
introduced in the literature. We have now rigorously computed the 
NLO contributions and analysed the obtained $K$-factors.

We define the acceptance window to coincide with the central rapidity
unit, $\Delta Y=\{(y,\phi): |y| \le 0.5,\, 0 \le \phi \le 2 \pi\}$,
$\phi$ being the azimuthal angle and $y$ the rapidity.  In a NLO hard
scattering of partons, we may have one, two, three or zero partons in
our rapidity acceptance.  All the partons are assumed massless, so the
transverse energy within $\Delta Y$ is the sum of the absolute values
$p_{Ti}$ of the transverse momenta of those partons whose rapidities
are in $\Delta Y$:
\begin{equation}
E_T = \epsilon(y_1)p_{T1} + \epsilon(y_2)p_{T2} + \epsilon(y_3)p_{T3},
\label{ET}
\end{equation}
where the step function $\epsilon(y_i)$ is defined as in \cite{EKL89},
\begin{equation}
\epsilon(y_i) \equiv \left\{
\begin{array}{ll}
        1 & \mbox{if $y_i\in\Delta Y$}\\
        0 & \mbox{otherwise.}
\end{array}
\right.
\end{equation}

In the LO, the transverse momenta are equal in magnitude:
$p_{T1}=p_{T2}=p_T$. The perturbative scatterings can in this case be
simply defined to be those with large enough transverse momentum,
$p_T\ge p_0\gg \Lambda_{\rm QCD}$, or $p_{T1}+p_{T2}\ge 2p_0$. This
generalizes to NLO as
\begin{equation}
p_{T1} + p_{T2} + p_{T3} \ge 2p_0,
\end{equation}
where $p_0$ is the {\em same} external parameter as in the LO case.
We emphasize that $p_0$ above is a {\em fixed} parameter which does
not depend on $\Delta Y$.

The IR safe measurement functions $S_2$ and $S_3$ \cite{KS92} can now
be written down. They are designed to answer the following question:
what is the amount of $E_T$ in $\Delta Y$ carried by the partons which
are produced in scatterings where at least an amount $2p_0$ of
transverse momentum is released?

We require both the definition of $E_T$ {\em and} the definition of
hard (perturbative) scatterings to be included in the measurement
functions, so for the $2\rightarrow 2$ scatterings we define
\begin{eqnarray}
S_2(p_1,p_2) =  
\Theta(p_{T1}+p_{T2}\ge 2p_0)
 \delta(E_T-[\epsilon(y_1)p_{T1}+\epsilon(y_2)p_{T2}])
\label{S2}
\end{eqnarray}
and  for the $2\rightarrow 3$ scatterings correspondingly
\begin{eqnarray}
S_3(p_1,p_2,p_3) =  
\Theta(p_{T1}+p_{T2}+p_{T3}\ge 2p_0)
\delta(E_T-[\epsilon(y_1)p_{T1}+\epsilon(y_2)p_{T2}+\epsilon(y_3)p_{T3}]).
\label{S3}
\end{eqnarray}

We emphasize that these measurement functions are not mere
generalizations of the jet cone $R$ of the conventional jet case
\cite{KS92,EKS89,EKS90,KUNSZT} to a rectangular rapidity window
$\Delta Y$ but they also carry information of the scale
$p_0$ which is not present in the computation for observable jets.
One should also keep in mind that contrary to the $E_T$ of single
inclusive jets, the initial $E_T$ we are considering here, is not a
directly measurable quantity, especially not for nuclear collisions at
RHIC and the LHC.  The question that we ask, however, is a
well-defined one, and can be answered by making an IR safe pQCD
computation analogous to that of observable jets \cite{KS92}.

Extending the definition of the semi-inclusive $E_T$-distribution of
\cite{EKL89} to NLO gives:
\begin{eqnarray}
\nonumber
\frac{d\sigma}{dE_T}\bigg|_{p_0,\Delta Y}&=&
\frac{d\sigma}{dE_T}\bigg|_{p_0,\Delta Y}^{2\rightarrow2} +
\frac{d\sigma}{dE_T}\bigg|_{p_0,\Delta Y}^{2\rightarrow3} \\ 
&=&
\frac{1}{2!}\int [d{\rm PS}]_2 
\frac{d\sigma^{\,2\rightarrow2}}{[d{\rm PS}]_2} S_2 + 
\hspace{3mm}\frac{1}{3!}\int [d{\rm PS}]_3 
\frac{d\sigma^{\,2\rightarrow3}}
{[d{\rm PS}]_3} S_3 
\label{dET}
\end{eqnarray}
in which momentum conservation correlates the transverse momenta as
${\bf p}_{T1}=-{\bf p}_{T2}$ in the $2\rightarrow2$ kinematics, and
${\bf p}_{T1}=-({\bf p}_{T2}+{\bf p}_{T3})$ in the $2\rightarrow3$
kinematics. The $[d{\rm PS}]_2$ and $[d{\rm PS}]_3$ stand for
$dp_{T2}dy_1dy_2$ and $dp_{T2}dp_{T3}d\phi _2 d\phi_3 dy_1 dy_2 dy_3$,
respectively.

The divergencies present in the partonic NLO cross sections can be
regulated by computing the squared matrix elements in $4-2\varepsilon$
dimensions, resulting in an explicit $\varepsilon^{-1}$ and
$\varepsilon^{-2}$ behaviour of the divergent terms. This was done
first by R.K. Ellis and Sexton \cite{ES86}. Using the $4-2\varepsilon$
dimensional squared matrix elements EKS have formulated the
calculation of observable cross sections for jet production
\cite{KS92,EKS89,EKS90,KUNSZT}. We will apply their approach, based on
the subtraction method, here.

As explained in \cite{KS92}, cancellation of the divergencies takes
place only if the three-parton measurement function $S_3$ reduces to
the two-parton one $S_2$ in soft and collinear limits.  Our
measurement functions in Eqs. (\ref{S2}) and (\ref{S3}) clearly fulfil
these criteria.  It is also worth mentioning that while the transverse
energy is a good quantity to compute, e.g. the number of gluons in
$\Delta Y$ would not make an IR safe measurement function without
specifying when two nearly collinear gluons are to be counted as one.

From Eq. (\ref{dET}) one obtains the first moment of the semi-inclusive 
$E_T$ distribution:
\begin{eqnarray}
\nonumber
\sigma\langle E_T\rangle_{p_0,\Delta Y} &\equiv&
\int_0^{\sqrt s} dE_T\,E_T \frac{d\sigma}{dE_T}\bigg|_{p_0,\Delta Y} \\
&=& \sigma \langle E_T\rangle_{p_0,\Delta Y}^{2\rightarrow2} 
+ \sigma \langle E_T\rangle_{p_0,\Delta Y}^{2\rightarrow3},
\label{sET}
\end{eqnarray}
where, after integrating the delta functions away in Eq.~(\ref{dET}), 
\begin{eqnarray}
\sigma \langle E_T\rangle_{p_0,\Delta Y}^{2\rightarrow2} =
\frac{1}{2!}\int [d{\rm PS}]_2
\frac{d\sigma^{\,2\rightarrow2}}{[d{\rm PS}]_2} \tilde S_2(p_1,p_2) 
\label{sET2}
\end{eqnarray}
and
\begin{eqnarray}
\sigma \langle E_T\rangle_{p_0, \Delta Y}^{2\rightarrow3} =
\frac{1}{3!}\int [d{\rm PS}]_3 \frac{d\sigma^{\,2\rightarrow3}}{[d{\rm PS}]_3} 
\tilde S_3(p_1,p_2,p_3).
\label{sET3}
\end{eqnarray}
The measurement functions for the first $E_T$-moment above are denoted by   
\begin{equation}
\tilde S_2(p_1,p_2) = \bigg[\epsilon(y_1)+\epsilon(y_2)\bigg]p_{T2}
\Theta(p_{T2}\ge p_0)
\end{equation}
and
\begin{eqnarray}
\tilde S_3(p_1,p_2,p_3) = 
\bigg[\epsilon(y_1)p_{T1}+\epsilon(y_2)p_{T2}+\epsilon(y_3)p_{T3}\bigg]
 \Theta(p_{T1}+p_{T2}+p_{T3}\ge 2p_0),
\end{eqnarray}
where $p_{T1}= |{\bf p}_{T2}+{\bf p}_{T3}|$. Naturally, also $\tilde
S_2$ and $\tilde S_3$ fulfil the IR criteria, which ensures that
$\sigma \langle E_T\rangle_{p_0, \Delta Y}$ is a well-defined IR safe
quantity to compute. From the general form of Eqs. (\ref{sET2}) and
(\ref{sET3}) we notice that, by replacing the measurement functions
$S_2$ and $S_3$ of Ref.  \cite{KS92} by $\tilde S_2$ and $\tilde S_3$
above, we can follow the formulation of the problem as given in
Ref. \cite{KS92}.  As all the relevant formulae with a complete
discussion of the cancellation of the several $\sim \varepsilon^{-1}$
and $\sim \varepsilon^{-2}$ singularities can be found in all detail
in \cite{KS92}, we now proceed directly to the numerical evaluation.

Our $\tilde S_2$ and $\tilde S_3$ are azimuthally symmetric, the
$\phi_2$-integrals giving a trivial factor $2\pi$, so basically only
three- and six-dimensional integrals remain to be done in $\sigma
\langle E_T\rangle_{p_0,\Delta Y}^{2\rightarrow2}$ and $\sigma \langle
E_T\rangle_{p_0,\Delta Y}^{2\rightarrow3}$, correspondingly.  For
$2\rightarrow2$ kinematics, the integration limits imposed by the
acceptance cuts in $\tilde S_2$ can be solved analytically, allowing
for a quick and accurate computation using a NAG library \cite{NAG}
subroutine based on an adaptive subdivision strategy.

Contrary to the $2\rightarrow2$ case, the kinematical cuts for the
three particle phase space are very complicated due to the additional
cuts introduced in the subtraction terms. To keep the counterparts of
the subtraction terms as they are given in \cite{KS92} (and which have
$2\rightarrow2$ kinematics), the kinematical cuts for the subtraction
terms in the $2\rightarrow3$ integrals must be strictly imposed
through the measurement functions $\tilde S_3$.  These integrals are
evaluated using a Monte Carlo subroutine of NAG.

Since the subtraction procedure adopted here is exactly that of
Refs. \cite{KS92,EKS89,EKS90} for production of inclusive high-$p_T$
jets, we are also able to make use of certain subroutines in the JET
program \cite{PROGRAM} of EKS.  In particular, we have used the
subroutines of JET for the functions in terms of which the squared
matrix elements were expressed in \cite{KS92}.  In addition, we have
adopted EKS's book-keeping method of the parton flavours, and
explicitly made use of their permutation tables for different
subprocesses.

For the NLO $\overline{\rm MS}$ parton distributions, we use the
GRV94-HO \cite{GRV94} as implemented in PDFLIB \cite{PDFLIB}. 
The renormalization scale and factorization scale are chosen equal, 
$\mu_R = \mu_F = \mu$. Our choice for the $2\rightarrow3$ terms is a 
generalization of the $2\rightarrow2$ case $\mu = N_{\mu}\times p_T$, 
\begin{equation}
\mu = N_{\mu}\times(p_{T1}+p_{T2}+p_{T3})/2
\label{scalechoice}
\end{equation}
where $N_{\mu}$ is a number of the order of unity.  The scale $\mu$ is
therefore IR safe in the same sense as the measurement functions are,
as required by the cancellation of the divergencies.

%%%%%%%%%%%%%%%%%%%%%%%%%%%%%%%%% FIGURE
\begin{figure}[ht]

\vspace*{-1cm}

\epsfysize=11cm
\centerline{\epsffile{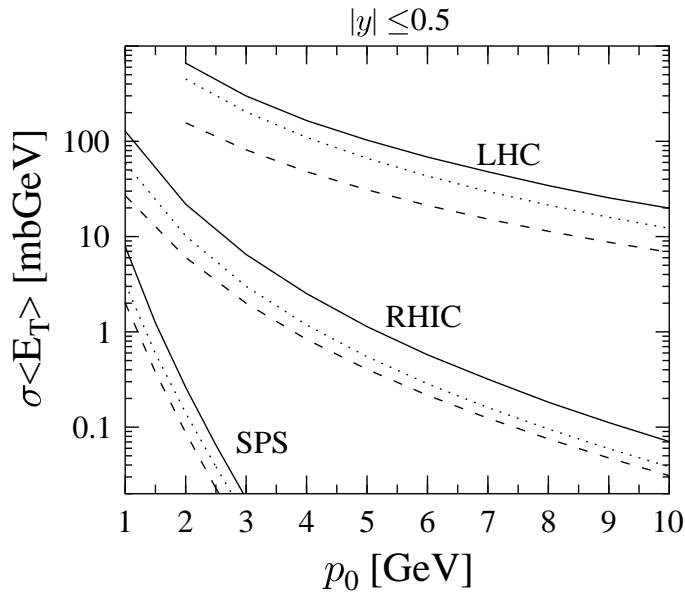}}

\vspace*{-1.5cm}

\caption[a]{\small The first $E_T$ moment $\sigma \langle
E_T\rangle_{p_0,\Delta Y}$ in the central rapidity unit $|y|\le0.5$
of the $E_T$ distribution (\ref{sET}). 
The results are shown for $pp$ collisions at $\sqrt s =5.5$ TeV (LHC),
200 GeV (RHIC) and 20 GeV (SPS) as functions of the parameter $p_0$.
The solid curves represent the full NLO calculation, the dotted ones
show the LO results computed with GRV94-LO parton distributions and
one-loop $\alpha_s$, and the dashed curves stand for the LO
computation with GRV94-HO parton distributions and two-loop
$\alpha_s$. All the scales are chosen (Eq.~\ref{scalechoice}) as
$\mu=(p_{T1}+p_{T2}+p_{T3})/2$.}
\label{sigmaET}
\end{figure}
%%%%%%%%%%%%%%%%%%%%%%%%%%%%%%%%%%%%

In Fig. \ref{sigmaET} we plot the first $E_T$ moment $\sigma \langle
E_T\rangle_{p_0,\Delta Y}$ in $pp$ collisions (i.e. no nuclear
shadowing is included) in the central rapidity unit $|y|\le 0.5$, as a
function of the minimum transverse momentum $p_0$ for $\sqrt s=5.5$
TeV, $200$ GeV and $20$ GeV.  Based on Ref. \cite{EKRT99},
for $AA$ collisions ($A\sim 200$) we expect the relevant range of
$p_0$ to be $p_0 \sim 2(1)$ GeV for the LHC (RHIC), so we show
$p_0=1...10$ GeV in the figure to see whether the NLO results become
very unstable relative to LO at the few-GeV transverse momenta.  The
difference between the two LO results at $\sqrt s=200$ GeV is mainly
due to the difference between the one- and two-loop $\alpha_s(\mu)$ used.
At $\sqrt s=5500$ GeV the differencies of the NLO and LO parton
distributions also become important. The $K$-factors, defined as
$K=$(full NLO)/LO, with LO in the denominator being either the dashed
or dotted curve, can now be directly read off from the figure: At
RHIC, as $p_0$ is decreased from $10$ to $1$ GeV, the $K$-factors
range from $2.3$ to $4.8$ and from $1.9$ to $2.2$, the latter being
the one evaluated against the truly LO result, i.e. the dotted
line. At the LHC the corresponding values are from $3.0$ to $4.2$ and
from $1.6$ to $1.5$ when $p_0$ is varied from 10 GeV down to 2 GeV.

%%%%%%%%%%%%%%%%%%%%%%%%%%%%%%%%% FIGURE
\begin{figure}[t]

\vspace*{-1.0cm}

\hspace{1cm}
\epsfysize=11cm
\centerline{\epsffile{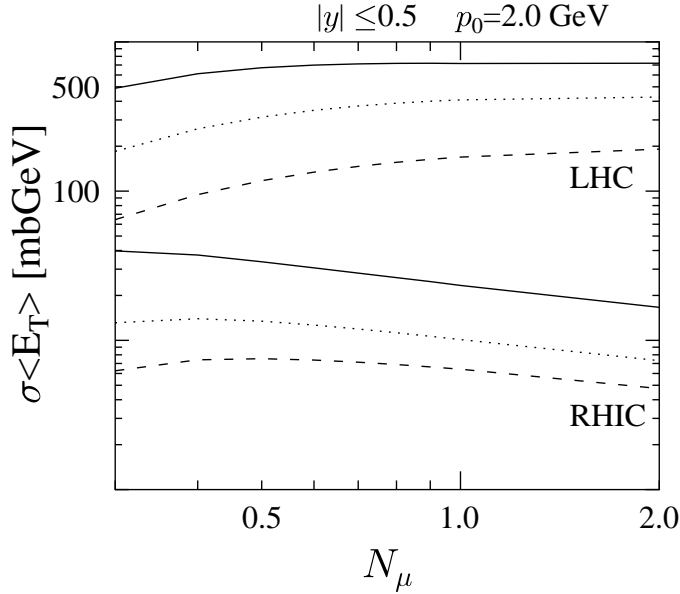}}

\vspace*{-1.5cm}

\caption[a]{ \small Scale dependence of $\sigma \langle
E_T\rangle_{p_0,\Delta Y}$ as a function of
$N_{\mu}=2\mu/(p_{T1}+p_{T2}+p_{T3})$ for the LHC (upper set
of three curves) and RHIC (the three lowest curves) energies at 
fixed $p_0=2$ GeV. Labelling of the curves is the same as in
Fig.~\ref{sigmaET}.}
\label{mudep}
\end{figure}
%%%%%%%%%%%%%%%%%%%%%%%%%%%%%%%%%%%%

In light of the inclusive one-jet production, \cite{EKS90}, the
$K$-factors are expected to depend on the scale choice $\mu$. In the
best case scenario from the viewpoint of applicability of pQCD at
$p_0\sim$ a few GeV, a region in $N_{\mu}\sim {\cal O}(1)$ would exist
where the NLO results would be stable against variations of the scale
choice, i.e.  independent of $N_{\mu}$ in Eq. (\ref{scalechoice}).
This is studied in Fig. \ref{mudep}, where the full NLO results and
the two LO calculations are shown as a function of $N_{\mu}$ for a
fixed value of $p_0=2$ GeV. The values of the curves at $N_{\mu}=1$
are the same as in the previous figure at $p_0=2$ GeV. At RHIC
energies, the scale dependence of the NLO-results is very similar to
the LO case.  For the LHC, the NLO results do {\em not} depend much on
the scale choice.  A similar behaviour is, however, found also in the
LO, so the stable behaviour cannot be attributed to a better
convergence of the perturbation series obtained by including the NLO
terms. The stable behaviour is due to the decrease of $\alpha_s(\mu)$
and the increase of $xg(x,\mu^2)$ with larger values of $N_{\mu}$. In
the $p_0$-region considered here, these effects virtually cancel. The
full NLO results seem therefore to be as (in)sensitive to the choice
of the scale as the LO results are both at the LHC and RHIC energies
at $p_0\sim$ a few GeV.

The following remarks are in order: First, consider the $K$-factor
calculated with respect to the dashed curve in Fig. 1. We note that
this ratio is quite stable, but presents a small increase towards
smaller values of $p_0$. This signals that we are working near the
borderline of perturbative QCD. On the other hand the $K$-factor
calculated against the truly LO result, shown by the dotted curve in
Fig. \ref{sigmaET}, does not increase, which implies that the NLO
results are stable all the way down to $p_0=1...2$ GeV.  Second, the
actual magnitude of the $K$-factors partly follows from the fact that
below the value $E_T=p_0$ there is a completely new contribution to
the minijet $E_T$ introduced by the NLO terms: the $E_T$-distribution
at $E_T<p_0$ is empty in LO but in NLO this is not the case anymore.
Therefore, 
more attention shoud be paid to the overall behaviour than to the 
absolute magnitude of the $K$-factors. The presence of the new kinematical
region in NLO allows us to conjecture that since this region would be
present also in the order next to NLO, the NNLO terms will be suppressed.  
Third, based on the jet
cone $R$ dependence of the inclusive jet production \cite{EKS90} one
might expect that the $K$-factors discussed above depend on the choice
for $\Delta Y$. However, as the LO contribution to $\sigma\langle
E_T\rangle_{p_0,\Delta Y}$ is almost linearly proportional to $\Delta
Y$, the $\Delta Y$ dependence of the $K$-factors should become much
weaker than in the inclusive one-jet case.

An attempt to compute $\sigma\langle E_T\rangle_{p_0,\Delta Y}$ in NLO
with the semi-hard region included has also been recently presented in
Ref. \cite{LO98}. The formulation \cite{LO98} is, however, a more
direct generalization of the inclusive jet production
\cite{KS92,EKS89,EKS90,KUNSZT}: $\Delta Y$ is introduced and $E_T$
within $\Delta Y$ is defined as we have done above but a cut-off $E_0$
in $E_T$ is introduced. Consequently, at $E_T<E_0$, the $E_T$
distribution is empty both in LO and especially in NLO. The
formulation \cite{LO98} thus misses the new, calculable, perturbative
contribution in this region.  In addition, a cut-off imposed in $E_T$
causes a $\Delta Y$ dependence of $p_0$ not present in our
formulation. We believe that to maximally account for the $E_T$
production from minijets, and thus to correctly generalize the
formulation of Ref. \cite{EKL89}, the starting point must be to first
fix the minimum overall transverse momentum $2p_0$ once and for all,
regardless whether the partons fall into $\Delta Y$ or not.  As
discussed above, this results in more than just an extension of the
conventional jet calculation.

We have in this letter extended the LO formalism
of Ref. \cite{EKL89} to NLO. Comparison against LO results, taken together
with the fact that a new kinematical region is contained in NLO, shows
that the NLO computation is a meaningful one, and should be taken into
account when evaluating initial transverse energy in nuclear collisions.

\bigskip\bigskip\noindent {\bf Acknowledgements:} We thank K. Kajantie and
V. Ruus\-kanen for discussions.  KJE is grateful to S. Ellis,
Z. Kunszt and D. Soper for several conversations regarding their NLO
jet-program. We also thank A. Leonidov for discussions regarding
Ref. \cite{LO98} and the Academy of Finland for financial support.
\vfill\eject


\begin{thebibliography}{50}

\bibitem{BM87}
        J.P.~Blaizot and A.H.~Mueller,
        Nucl. Phys. B289 (1987) 847.

\bibitem{KLL87}
        K. Kajantie, P.V. Landshoff and J. Lindfors, 
        Phys. Rev. Lett. 59 (1987) 2527.

\bibitem{EKL89}
        K.J.~Eskola, K.~Kajantie and J.~Lindfors,
        Nucl. Phys. B323 (1989) 37.

\bibitem{EKRT99}
        K.J.~Eskola, K.~Kajantie, P.V. Ruuskanen and K. Tuominen, 
        Preprint JYFL-8-99, hep-ph/9909456, to appear in Nucl. Phys. B.

\bibitem{KS92}
        Z. Kunszt and D.E. Soper, Phys. Rev. D46 (1992) 192.

\bibitem{EKS89}
        S.D. Ellis, Z. Kunszt and D.E. Soper, Phys. Rev. D40 (1989) 2188.

\bibitem{EKS90}
        S.D. Ellis, Z. Kunszt and D.E. Soper, Phys. Rev. Lett. 64 (1990) 2121.

\bibitem{KUNSZT}
        Z. Kunszt, ETH-TH-96/05, 
        lectures presented at TASI 95, Boulder, June 1995; hep-ph/9603235.

\bibitem{LO98}
         A. Leonidov and D. Ostrovsky, FIAN-TD-24-98, hep-ph/9811417.

\bibitem{ES86} 
        R.K. Ellis and J.C. Sexton, Nucl. Phys. B269 (1986) 445.

\bibitem{NAG}
	NAG Fortran Library, Mark 18.

\bibitem{PROGRAM}
        S.D Ellis, Z. Kunszt and D. E. Soper, 
        {\em JET} version 3.4, 18 March 1997.
        
\bibitem{GRV94}
        M. Gl\"uck, E. Reya and A. Vogt, Z.Phys. C67 (1995) 433.

\bibitem{PDFLIB}
        H. Plothow-Besch, 
        PDFLIB Version 7.09, W5051 PDFLIB, 1997.07.02, CERN-PPE.

\end{thebibliography}
\end{document}